# Inconsistent illusory motion in predictive coding deep neural networks

**Authors and affiliations:**
O.R. Kirubeswaran[1] & Katherine R. Storrs[2,3,4]
1. Indian Institute of Science Education and Research Pune
2. Department of Experimental Psychology, Justus Liebig University Giessen
3. Centre for Mind, Brain and Behaviour (CMBB), University of Marburg and Justus Liebig University Giessen
4. School of Psychology, University of Auckland

**Contact details for corresponding author:**
O.R. Kirubeswaran,
Department of Biology, Indian Institute of Science Education and Research Pune, Pune, 411008, Maharashtra, India.

email: kirubes.waran@students.iiserpune.ac.in

## Abstract:

Why do we perceive illusory motion in some static images? Several accounts have been proposed based on eye movements, response latencies to different image elements, or interactions between image patterns and motion energy detectors. Recently, PredNet, a recurrent deep neural network (DNN) based on predictive coding principles was reported to reproduce the "Rotating Snakes" illusion, suggesting a role for predictive coding in illusory motion. We begin by replicating this finding, and then use a series of "*in silico* psychophysics" experiments to examine whether PredNet behaves consistently with human observers for greyscale, gradient-based, and simplified variants of the Rotating Snakes and Fraser-Wilcox illusion stimuli. We also measure response latencies to individual elements of the Rotating Snakes pattern by probing internal units in the network. A pretrained PredNet model predicted illusory motion for all subcomponents of the Rotating Snakes stimulus, consistent with human observers. However, we found no simple response delays in internal units, unlike evidence from physiological data. PredNet's detection of motion in gradients seemed dependent on contrast, whereas human perception of motion in gradients depends predominantly on luminance. Finally, we examined the robustness of the illusion in a set of ten PredNets of identical architecture, retrained on the same video data. There was large variation across network instances in whether they reproduced the Rotating Snakes illusion, and what motion, if any, they predicted for simplified variants of the illusion. Unlike human observers, no network predicted illusory motion for greyscale variants of the Rotating Snakes pattern. Our results sound a cautionary note: even when a DNN successfully reproduces some idiosyncrasy of human vision, more detailed investigation can reveal inconsistencies between humans and the network, and between different instances of the same network. Given the inconsistency of the Rotating Snakes illusion in PredNets trained from different random initializations, we conclude that predictive coding does not reliably give rise to human-like illusory motion.

## Keywords:

Visual illusions; motion perception; deep neural networks; predictive coding; peripheral drift illusion.

# Introduction:

Our visual processing is exquisitely adapted to seeing things moving in the world—yet certain static images can also create vivid sensations of motion where there is none. Peripheral drift illusions[1][2] are a class of static patterns that can invoke powerful impressions of motion, especially when seen in the visual periphery, and may help expose the mechanisms underlying motion perception. The Rotating Snakes Illusion is an example in which a strong illusory percept of rotation is induced by a pattern of periodically repeating colour elements: white, yellow, black, and blue (the illusion also works in a monochromatic version, with only luminance changes[3]). The direction of illusory motion is perceived in the aforementioned order. The Rotating Snakes Illusion has attracted considerable attention from the vision science community[4][5][6][7][8][9] as well as from the general public due to the robust nature of the illusion in humans as well as other animals[10][11][12][13][14].

There are several proposed explanations of how the Rotating Snakes Illusion works. Some theories accredit the perception of peripheral drift illusions to fixational eye movements such as microsaccades or other oculomotor functions like blinks and drifts[8][5][15][16]. Certainly, the Rotating Snakes Illusion seems to require the stimulus to be transient in some way, and illusory rotation appears strongest when saccadic eye movements are made while viewing the stimulus[17]. But it can also be perceived when the image is briefly flashed[18] or moved between locations while the observer fixates (see demos in Conway et al (2005) and Backus & Oruc (2005))[16][8][5][15]. Several mechanistic explanations of the illusion are based on luminance or contrast processing latency differences across the pattern. If elements in the Rotating Snakes pattern that have higher luminance or contrast are processed earlier than the lower luminance or contrast elements, these could elicit local motion signals in the brain[18][1][4], and the integration of many such local signals across the circularly repeating elements of the pattern could give rise to coherent rotational motion. Most recently, Bach & Atala-Gérard (2020)[19] showed computationally that the Rotating Snakes pattern gives rise to consistent directional motion when input to motion energy detectors with a saturating nonlinear output function, which are widely thought to form the basis of cortical motion perception[20].

Watanabe et al. (2018)[21] reported that the Rotating Snakes illusion is also perceived by a Deep Neural Network (DNN) built on predictive coding principles, called PredNet[22][21]

(Figure 1). PredNet is an artificial recurrent DNN built to implement hierarchical predictive coding[22]. It uses prediction as an unsupervised learning objective, learning to predict future frames of video sequences, and thereby developing internal features that allow it to better represent the complex input. Predictive coding theory posits that the brain constantly predicts the incoming sensory input from the world and tries to minimize its prediction error, i.e the difference between the actual input and its prediction. The theory suggests that each region in the cortical hierarchy makes local predictions in a top-down fashion and computes prediction errors by comparing those predictions to their subsequent input from preceding regions. These prediction errors are then fed to higher cortical regions in a bottom-up manner, giving rise to a recurrent process through which the brain learns to construct and update an increasingly rich internal model of its environment. This view of cortical processing as a hierarchical generative model that predicts sensory inputs has received widespread interest as a potentially unifying theory of the underlying mechanisms in visual processing and beyond [23][24][25][26][27][28].

Watanabe et al., 2018[21] trained PredNet on real-world video sequences from the First-Person Social Interactions Dataset[29], a video dataset of head-mounted camera footage (see example frames in Figure 1B), then presented it with the Rotating Snakes pattern as static frames. They report that the network, when prompted to predict future frames, extrapolates the pattern as rotating in the same direction human observers see illusory motion (and does not predict any motion for a closely matched non-illusion control image). The authors suggest that since PredNet appears to be susceptible to the Rotating Snakes illusion, predictive coding principles could explain the causal mechanisms underlying this and other peripheral drift illusions[30]. Specifically, they propose that PredNet's prediction of movement in the Rotating Snakes illusion implies it has converged on similar motion processing mechanisms as found in the human visual system. Such a claim is not necessarily incompatible with more traditional mechanistic accounts of the illusion, based on response latencies[4][18] or motion energy detectors[19]. Rather, underlying constraints in the brain such as the need for energy-efficient coding could give rise to predictive processes[31][28], which in turn determine the properties of visually responsive neurons.

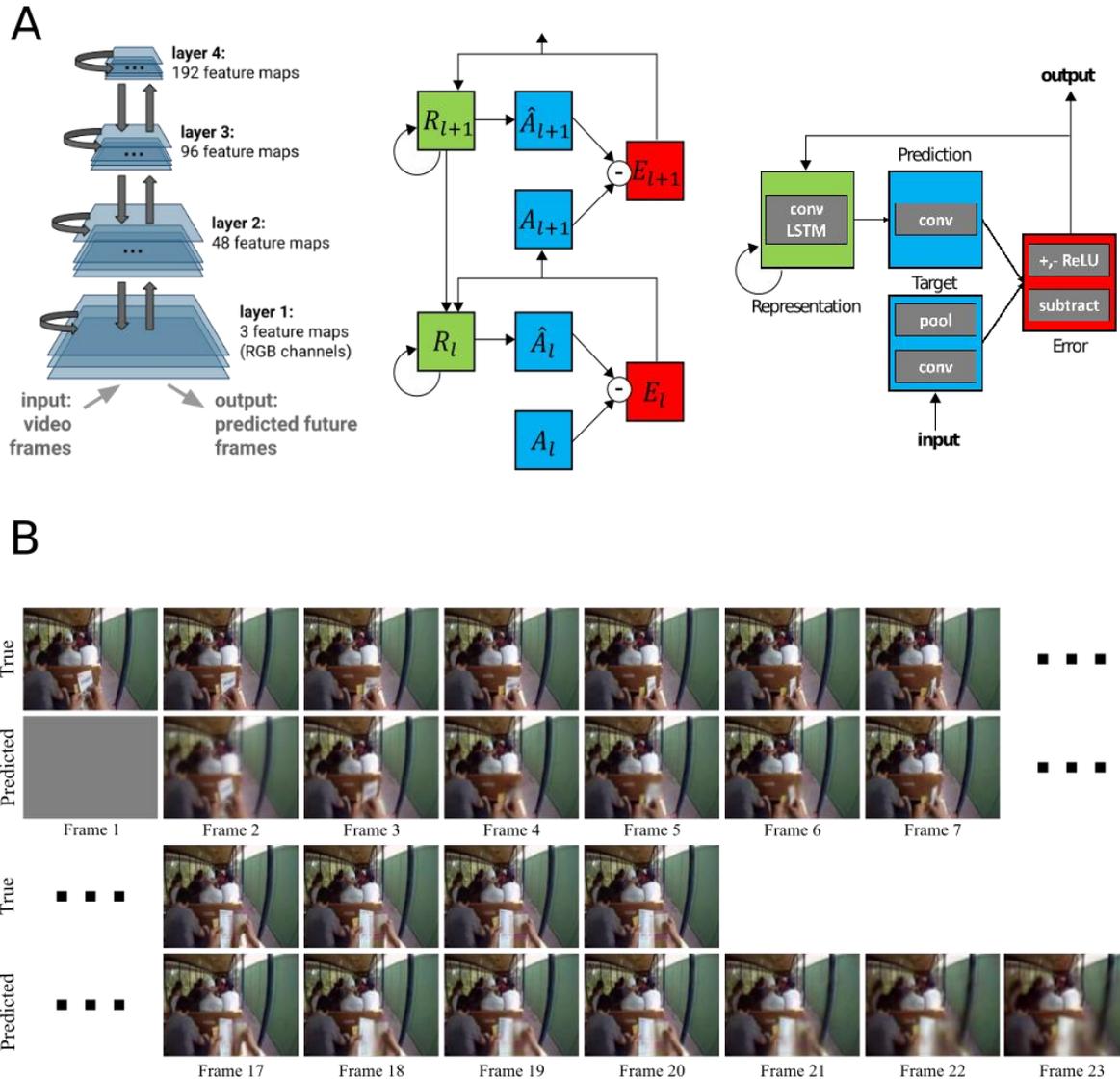

*Figure 1: PredNet's architecture and example future frame predictions. (A) Schematic illustration of the full architecture of the four-layer PredNet recurrent convolutional neural network (left) and more detailed schematics of the processing steps in two layers (middle) and within a single layer (right). Middle and left panels are reproduced with permission from Lotter, Kreiman & Cox (2017). (B) True input frames (top rows) and predicted next frames (bottom rows) performed by PredNet on an example video sequence not seen by the network during training. Frames 21-23 are extrapolated predictions beyond the 20$^{th}$ input frame. Input frames are reproduced from the First-Person Social Interactions Dataset: http://ai.stanford.edu/~alireza/Disney/.*

In order to evaluate the strength of evidence for predictive coding's role in peripheral drift illusions, we examine how PredNet perceives illusory motion in more detail, using a range of psychophysical stimuli that have helped shed light on the illusions' mechanisms in humans. We present the network with additional peripheral drift illusion stimuli (full, non-illusion, and greyscale versions of the Rotating Snakes Illusion, and a colour-based variant of the Fraser-Wilcox illusion[32]), and with stripped-down versions of these (bar-pair elements constituting the Rotating Snakes Illusion, and gradients comprising the Fraser-Wilcox illusion). We also probe the internal units of PredNet to see if they obey the contrast (or) luminance response properties described by Conway et al., 2005[18] and Faubert & Herbert, 1999[1]. This would provide evidence both that PredNet shares similar mechanisms with biological motion processing, and that these tuning properties of motion-sensitive neurons might indeed arise from predictive coding principles[33]. Further, we perform all analyses on both a pretrained network shared by Watanabe et al (2018)[21], as well as on new network instances that we trained on the same data (ten PredNets trained on colour videos and ten trained on greyscale versions of the same videos). Retraining deep neural network architectures from different random initial states can lead to drastic differences in performance and internal representations learned by the network[33]. The robustness of the illusion across network instances is fundamental to the claim that predictive coding principles give rise to peripheral drift effects.

The paper is structured as follows: We first set out to replicate, in Experiment 1, the basic observation that the Rotating Snakes illusion is predicted by the PredNet networked trained and shared by the research group of Watanabe et al (2018). Having done this, Experiments 2-4 then more deeply investigate consistency between the model and previously reported human data, in three *in silico* "psychophysical" and "electrophysiological" experiments using reduced versions of peripheral drift illusion stimuli. Finally, Experiment 5 asks whether illusory motion phenomena are robustly found in networks of this type, by retraining new networks of the same architecture from different random initial states.

# 1  Experiment 1: Reproducing peripheral drift illusions in the pretrained PredNet

We first tested whether we could reproduce the observation of the Rotating Snakes Illusion and its variants in a PredNet network trained and shared by Watanabe et al. (2018)[21].

## 1.1  Methods:

### 1.1.1  Network and training

The Deep Neural Network (DNN) architecture used in this study is a PredNet[22] network implemented in PyTorch (https://github.com/eijwat/prednet_in_pytorch.git). The PredNet configuration used, shown in Figure 1, consists of 4 hierarchically stacked layers. Each layer computes a prediction of its input at the next time point, and this is compared against the actual next input to calculate error signals. These error signals are then propagated up the hierarchy to the subsequent layer. Each layer (n=1-4) is composed of four modules containing different types of unit: input convolution units ($A_n$), recurrent representation units ($R_n$), prediction units ($\hat{A}_n$), and error units ($E_n$). The representation units are convolutional long short-term memory (ConvLSTM) units[34][35]. The number of channels for the representation units are 3, 48, 96, and 198 for layers 1 to 4 respectively, all with the same kernel size (3x3). In each layer, the representation units integrate information from errors, subsequent layers, and their own past states, to generate predictions ($\hat{A}_n$) of what their own input ($A_n$) will be at the next time step. The outputs of layer 1 representation units are pixel-wise predictions of what the input video sequence will look like at the next frame. In each layer, the prediction ($\hat{A}_n$) is subtracted from the input ($A_n$) to update the error units ($E_n$, with positive and negative errors expressed in separate sets of channels), which become the input of the next layer of the network after a convolution and a max-pooling operation (2x2). At the beginning of each input video sequence, activations in both the representation and error units are initialized to zero.

The model was trained on 20-frame sequences of 120x160 pixel RGB videos from the FPSI (First Person Social Interaction) dataset[29]. This dataset consists of footage from 6 subjects walking around with cameras mounted on top of their caps, proceeding with their day at the Disney World Resort in Orlando, Florida. We used the pretrained model[1] from Watanabe et al (2018), by downloading the weights shared by those authors here: https://figshare.com/articles/dataset/Sample_Weight_Models_for_PredNet_in_Pytorch_Rotating_Snake_Illusions_/16572065/1 (final training checkpoint "fpsi136_410000"). The model was trained on 410,000 frames of images from the 29v dataset which is a concatenated version of all the individual FPSI data.

### 1.1.2 Stimuli

We presented the pretrained PredNet network with two colour versions of the Rotating Snakes image, as used by Watanabe et al. (2018)[21]: one with asymmetrically arranged elements that induces counter-clockwise motion in human observers, and one with symmetric elements that does not induce illusory motion (Figure 2A, first and third image). We also presented the network with a greyscale version of the standard Rotating Snakes Illusion image, which is an effective illusory motion stimulus for humans (Figure 2A, second image). Many peripheral drift illusions, including the original Fraser-Wilcox Illusion, involve smooth gradients rather than the asymmetric discrete elements that make up the Rotating Snakes Illusion. We therefore also presented the network with a colour-based variant of the clockwise Fraser-Wilcox Illusion devised by (Kitaoka, 2014), available here: http://www.psy.ritsumei.ac.jp/~akitaoka/OFWtypeVe.html (Figure 2A, rightmost panel).

---

[1] Note that this pretrained PredNet is a more recent model than that provided with the original Watanabe et al (2018) publication. The more recent model corrects an error in the training data of the initially shared network (personal communication, Eiji Watanabe, 6th September 2021).

### 1.1.3 Quantifying predicted motion

To probe illusory motion for static images in PredNet we followed the same procedure as Watanabe et al. (2018): test images were input to the network as a video sequence of 20 static frames, each of which is a 120x160 pixels image in the PNG format with colour encoded in RGB coordinate space. The PredNet generates three extrapolated predictions, each using the previous frame as their input. For instance, the $21^{st}$ frame is generated using the $20^{th}$ predicted frame as its input, the $22^{nd}$ frame is generated using the $21^{st}$ extrapolated frame as its input, and so on. We measured the strength and direction of motion between the $20^{th}$ and the $21^{st}$ predicted frames using the Farneback optic flow algorithm[36] implemented in the OpenCV package for Python (window size = 30). The Farneback algorithm is a type of dense optical flow algorithm that calculates the motion vectors for all the pixels between any two frames.

### 1.2 Results:

The pretrained PredNet from Watanabe et al (2018) (Figure 2B, bottom panel) predicted counter-clockwise rotational motion when shown the static coloured Rotating Snakes Illusion, consistent with human perception. It did not predict any rotational motion in a closely matched "non-illusion" version of the Rotating Snakes image, in which elements are arranged symmetrically and humans perceive no peripheral drift (Figure 2B) [18]The model also predicted strong clockwise rotational motion for a version of the Fraser-Wilcox Illusion that employs colour gradients, again consistent with human observers (Figure 2B, rightmost panel). However, unlike human observers, the pretrained PredNet predicted no rotational motion for a grayscale version of the Rotating Snakes image[18] (Figure 2B, second panel). This is a surprising finding and challenges the interpretation of PredNet as a comprehensive model of human peripheral drift perception. In a later section we investigate whether training new

PredNet models on grayscale video data leads to networks that capture grayscale illusory motion, and find it does not (see Section 5.2.1 "Reproducibility of peripheral drift illusions in PredNet networks" and Supplementary Figure 1). Nevertheless, the pretrained model presented by Watanabe et al (2018) does indeed capture the stronger or more commonly observed coloured version of the Rotating Snake illusion, as well as another colour-based peripheral drift illusion not tested by those authors, so we proceeded to investigate it further.

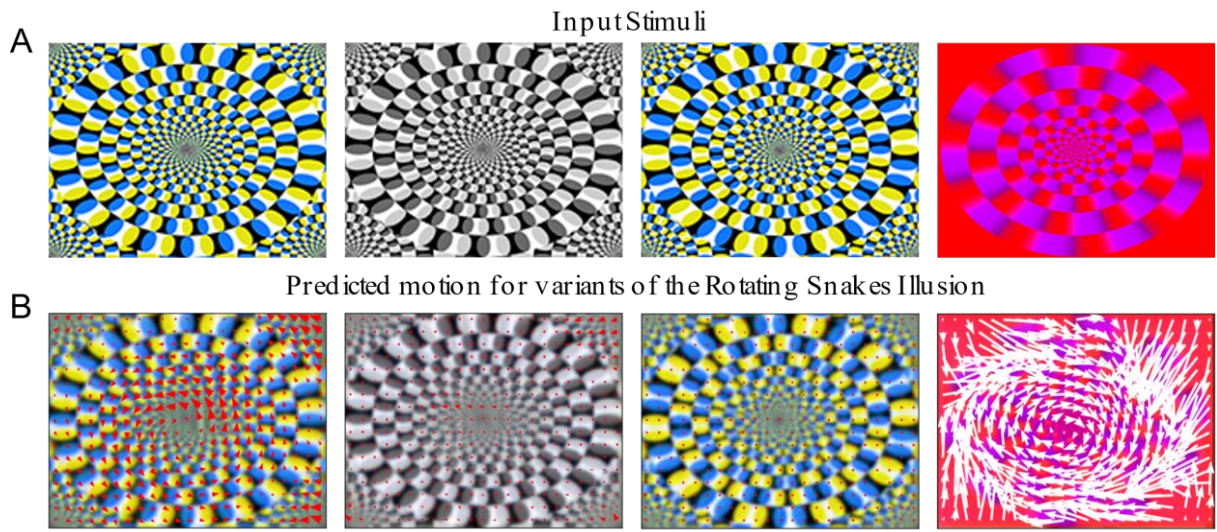

*Figure 2: Motion predicted by the pretrained PredNet for peripheral drift illusion stimuli. (A) The four static images input to the network as test stimuli; from left to right: a counter-clockwise Rotating Snakes Illusion stimulus; a greyscale version of the same image; a symmetric version of the same pattern, which does not induce illusory motion in human observers; and a colour-based variant of the clockwise Fraser-Wilcox Illusion. All four input images are copyright Akiyoshi Kitaoka. (B) Motion predicted in each test stimulus by the PredNet model pretrained by Watanabe et al. (2018). Arrows depict optical flow vectors detected between the 21st and 22nd extrapolated frames after the network has received 20 identical static frames of the respective image. [Figure should be viewed in colour.]*

## 2    Experiment 2: Examining predicted motion for the subcomponents of the Rotating Snakes Illusion

Having established that colour-defined peripheral drift illusions could be successfully reproduced in the original pretrained PredNet network, we went on to test different possible mechanisms underlying the Rotating Snakes Illusion. We did this by first examining the magnitudes and directions of motion arising from different pairs of elements in the Rotating Snakes Illusion. Different accounts of the Rotating Snakes Illusion predict different amounts or directions of motion arising from different pairs of elements within the full Rotating Snakes image.

A luminance-dependent processing latency predicts that the areas of higher luminance are processed faster in the visual system than the areas with lower luminance[1][37][38]. Thus, in a grayscale version of the rotating snakes illusion, this manifests as local motion signals that arise along the direction of white–light gray or dark gray–black element pairs as a consequence of different processing latencies of luminance information. Alternatively, a contrast-dependent processing latency suggests that the motion detectors in the higher areas of the brain whose receptive fields are large enough to span the element pairs of the Rotating Snakes Illusion process the higher contrast bar earlier than the lower contrast one[18][39]. However, simple luminance- or contrast-based latency delays predict that some elements within the rotating snakes stimulus should appear to move in the opposite direction to the overall perceived direction of the illusion. Bach & Atala-Gérard's (2020)[19] explanation based on nonlinearly-saturating motion energy detectors also predicts that nearby pairs of elements may appear to move in opposite directions, depending on their relative luminance. Conway et al., (2005)[18] reports psychophysical evidence that all element pairs in fact induce illusory motion in the same direction, and proposed an explanation combining a contrast-dependence processing latency[40] with "reverse phi" apparent motion[41][42][43].

## 2.1 Methods:

### 2.1.1 Stimuli

To test the prediction in our networks we created minimalist versions of the Rotating Snakes Illusion, comprising just two bars on a grey background. The input stimulus is a 120x160 pixel image in which the two bars occupy 30x10 pixels each at the centre of the image. Four such stimuli were created, each containing pairs of successive elements in the colour version of the rotating snakes illusion: black–blue, blue–white, white–yellow, yellow–black. Similarly, grayscale versions consisting of black–dark grey, dark gray–white, white–light gray, light gray–black elements were also created. Note that no illusory motion is perceived upon casual viewing of these bar-pair stimuli. However, the use of such stimuli is inspired by Conway et al.[18]'s study which reports subtle perceived motion for repeating patterns of such element pairs in a psychophysical experiment.

### 2.1.2 Measuring direction and speed of predicted motion

Each stimulus was input to the network as a sequence of 20 static video frames, and we measured motion between the 20th and 21st predicted frames. We quantified the direction and strength of predicted motion for each of the element pairs comprising the Rotating Snakes Illusion in both the colour and greyscale versions using optical flow analysis as described in section 1.1.3. The optical flow analysis between the network-predicted frames detected motion vectors of various directions and magnitudes at every pixel (see Figure 3, left panels, where the motion vector obtained at every $7^{th}$ pixel is plotted for visualisation purposes). Since there was substantial variation in optical flow magnitude and direction across the image, we computed the bootstrapped mean horizontal motion over all detected optical flow vectors (1000 samples, with replacement) for each stimulus image. For all stimuli, human perception predicts rightwards motion (positive horizontal motion values)[18].

## 2.2 Results:

The pretrained PredNet predicts rightward motion for most subcomponents of both the colour as well as the grayscale version of the Rotating Snakes Illusion, and no subcomponents gave rise to consistent motion in the direction opposite to the overall illusion. This is broadly in agreement with the perceptual findings of Conway et al. (2005) and their reverse-phi contrast-latency explanation, which suggests that motion should be perceived rightwards for all subcomponents of the Rotating Snakes Illusion[3]. The colour stimuli produced more robust illusory motion than the greyscale stimuli (Figure 3B) for this pretrained PredNet model, with significant motion predicted for three of the four colour stimuli, and two of the grayscale stimuli, always in the rightwards direction (two-tailed bootstrap confidence interval of 99.5%).

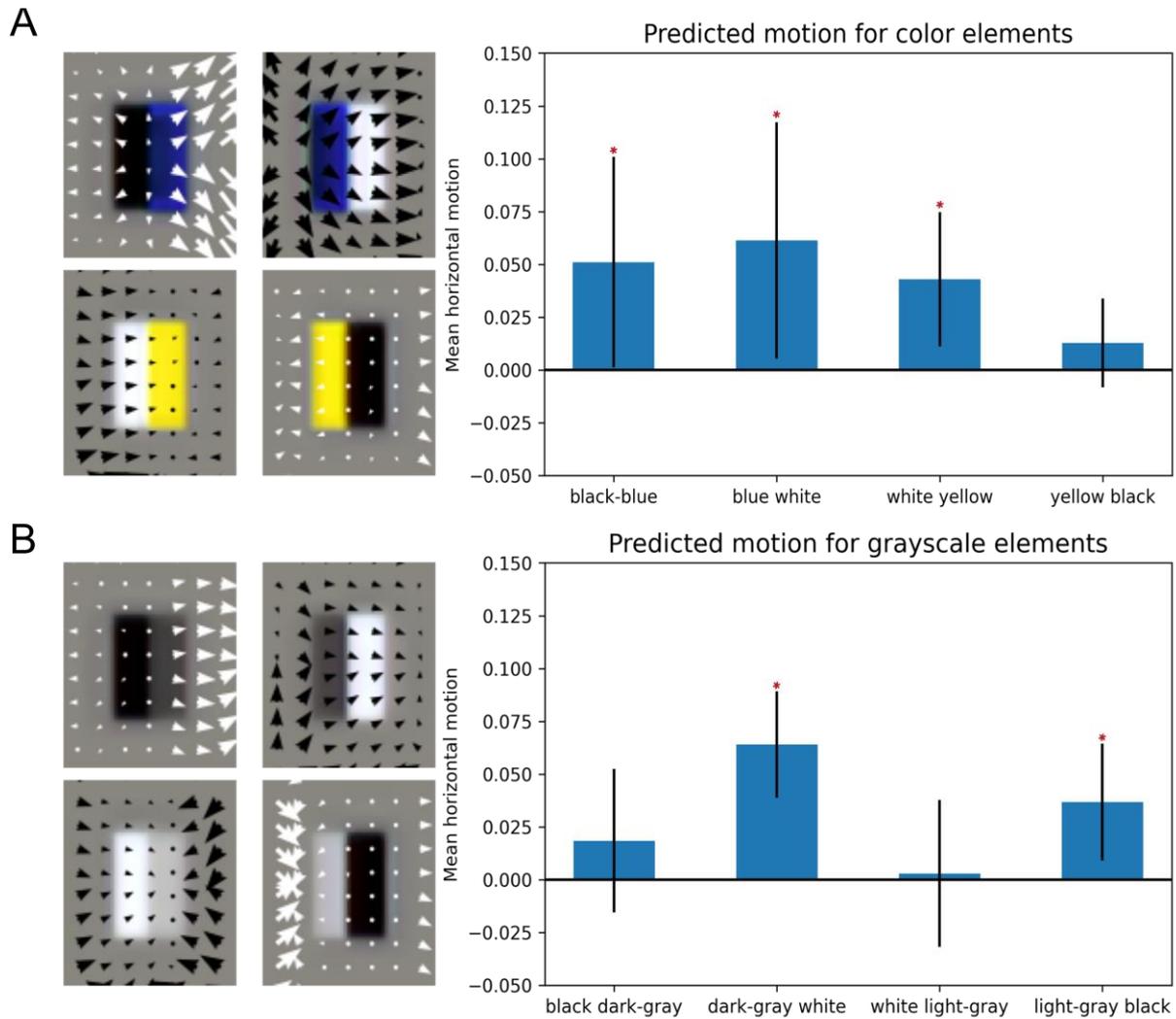

*Figure 3: Predicted motion for subcomponents of the Rotating Snakes Illusion in the pretrained PredNet.* (A) A colour and (B) a grayscale version of element pairs that make up the rotating snakes illusion are presented to PedNet. Left panels show the optical flow vectors detected in the element pair stimuli by the original network pretrained by Watanabe et al. (2018). The arrows indicate the direction and magnitude of optical flow calculated between the 20th and 21st frame predicted by PredNet after being presented with a static element pair image for 20 frames. To the right, barplots show the average direction and magnitude of horizontal motion predicted across all image locations for each stimulus (bar groups). The height of the bars represents the mean horizontal motion detected by the model, where a positive sign indicates rightward motion. Error bars indicate the 99.5% confidence interval of 1000 bootstrap samples of the mean, and red asterisks above bars indicate significant predicted motion.[Figure should be viewed in colour.]

## 3 Experiment 3: Measuring response latency in the network

Several theories of illusory motion suggest that the perception of such motion depends on differences in neural response latencies to stimuli based on differences in either luminance or

contrast[18][3][1][4][44]. Conway et al. (2005) found that contrast-dependent neural response latencies in macaque's direction-selective cells in V1/MT to individual elements of the rotating snakes illusion could predict the direction of the perceived illusory motion[18]. We therefore assessed whether a similar explanation might be true in PredNet by measuring response latencies to the sudden onset of individual bars that constitute the illusion. If activity in the network changes more quickly in response to the onset of some elements than others, this could mimic the temporal activation profile of a genuinely moving stimulus, and cause the network to incorrectly extrapolate motion.

## 3.1 Methods:

To test this, we created sequences of ten blank grey frames, followed by ten frames of a single bar shown at the centre of the image. The image and bar had the same dimensions as in the bar-pair stimuli used in Experiment 2. We created three such sequences, using either a white, blue, or light grey bar, because [18]these were the elements of the pairs that had created the largest (blue and white) and smallest (white and light grey) predicted motion in Experiment 2. We recorded the network's internal activation to each frame of these "sudden onset" sequences. Specifically, we recorded activations of the representation units ($R_l$) and error units ($E_l$) in the first layer for each frame. We then averaged over the spatial locations in the image (i.e., mean over convolutional units) to obtain an average activation in each feature map at each frame.

## 3.2 Results:

Activation responses in all three feature maps of both the representation units ($R1$; Figure 4B, D) and error units ($E1$; Figure C, E) appear to change immediately at the onset of the bar stimulus for all bar colours. Unlike the contrast-dependent latency responses found in the directionally selective cells of the primary visual cortex of the macaque monkeys, there doesn't appear to be a simple response delay between stimuli of different contrasts and colours within PredNet. However, the time courses of responses to blue vs. white bars (which induce strong predicted motion when paired; Figure B, C) are more different from one another than are the time courses of responses to light grey vs. white bars (where little or no motion is predicted; Figure D, E). Therefore, although we don't find evidence for a simple contrast-dependent processing delay, it remains possible that there are other more complex differences in temporal responses to elements related to this illusion.

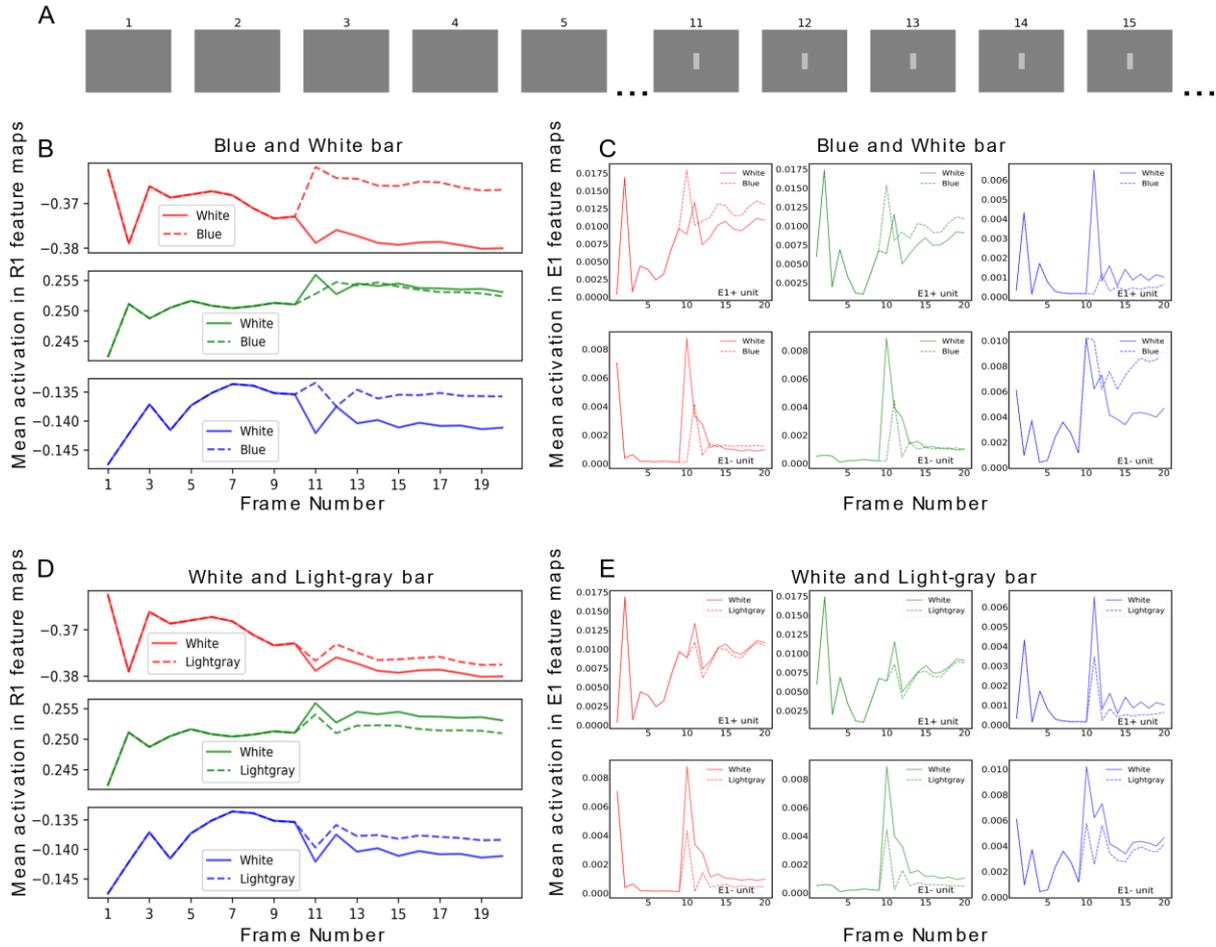

*Figure 4: Response latencies within the pretrained PredNet.* (A) An example test stimulus used to measure the latency of the responses to constituent bars of the illusion. The sequence transitions from an empty 120x160 pixel image for the first ten frames to an image containing one constituent bar for the subsequent ten frames. (B) and (D) show the average activation of the representation units (R1) in the first layer of the original pretrained PredNet within each feature map (corresponding to the red, green, and blue image channels, as indicated by line colour) for each frame. (C) and (E) show the average activation of the error units (E1) in the first layer of the PredNet within each feature map (positive (top row) and negative (bottom row) error populations of red, green, and blue feature maps) for each frame. The solid line in all plots shows the timecourse of responses when a white bar appears after ten blank frames. The dashed line in (B) and (C) shows the timecourse of responses to a blue bar (white–blue element pairs elicited the most predicted illusory motion in Experiment 2), while the dashed line in (D) and (E) shows responses to a light-grey bar (white–light grey element pairs elicited the least predicted illusory motion in Experiment 2).

## 4    Experiment 4: Effect of contrast and luminance on predicted motion

Finally, we tested the predictions of the network on gradient stimuli with different backgrounds. Placing the same gradient on different backgrounds creates regions (bar ends) that have high/low contrast with respect to the background. This allows us to discern the effect of luminance vs. contrast differences on the predicted motion. A study by von Grünau et al. suggests that for a bar filled with a light–dark gradient, motion is always perceived towards the higher luminance end of the bar by humans regardless of whether the gradient bar is presented on a light or dark background[37]. Thus, they conclude that the contrast of a simple gradient stimulus relative to its background plays no role in the perception of illusory motion (although, for the more complex patterns constituting the Rotating Snakes illusion, direction of perceived motion depends non-uniformly on the luminance of elements[45]. Thus, we tested similar gradient stimuli in the network, dissociating luminance from contrast by changing the background on which they were presented. Because the network generally predicted stronger motion for coloured stimuli than greyscale stimuli, we used the component colours of the original rotating snakes illusion to construct our stimuli.

### 4.1 Methods:

We created four stimuli: blue–black gradient bars on either a blue or black background and yellow–white gradient bars on either a yellow or white background. The input stimulus is a 120x160 pixel image in which the gradient bar occupies 20x80 pixels at the center of the image. Each stimulus was input to the network as a sequence of 20 static video frames, and we measured motion between the 20th and 21st predicted frames using an optical flow analysis, then computed the bootstrap mean horizontal motion over the detected optical flow vectors, as described in Experiment 2 Methods.

### 4.2 Results:

Predicted motion for blue–black gradients is much stronger than for yellow–white gradient stimuli (Figure 5A). However, the pretrained PredNet always predicted motion towards the location of the higher contrast end in the gradient stimuli, regardless of where the highest luminance location lay (Figure 5B). This is unlike the human psychophysical effect reported by von Grünau et al., where perceived motion moved towards the lower-luminance end of such gradient stimuli, regardless of contrast[37].

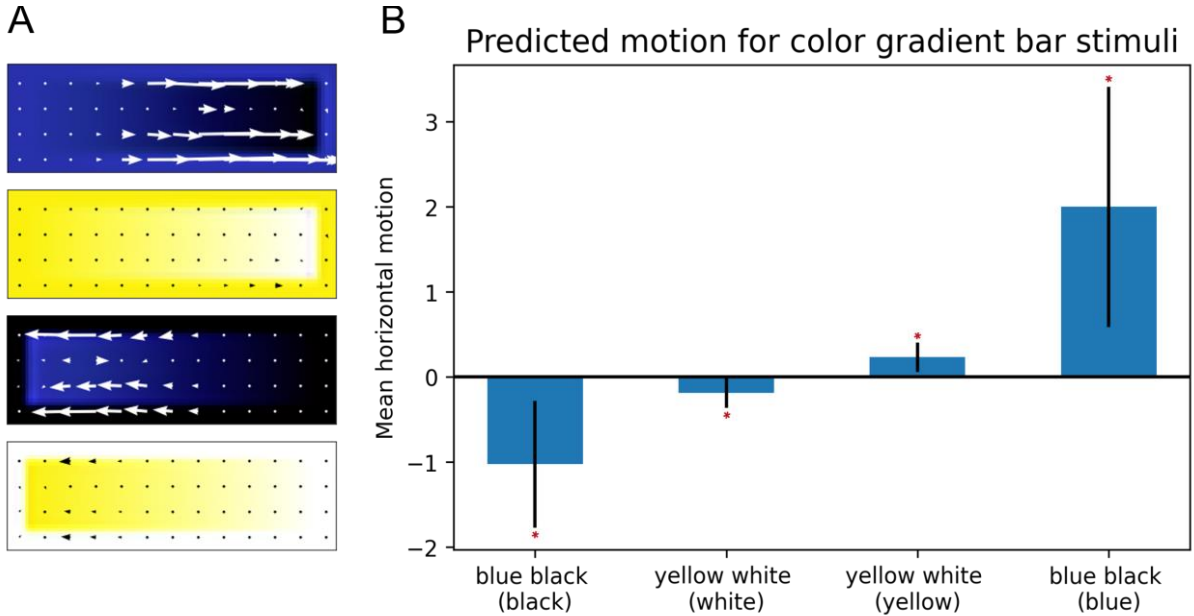

*Figure 5: **Predicted motion for gradient bars in the pretrained PredNet.** Colour gradient bars similar to the stimuli used by von Grünau et al, 1995 are presented to PredNet. (A) Optical flow vectors detected in the colour gradient bars by the original network pretrained by Watanabe et al. (2018). The arrows indicate the direction and magnitude of optical flow calculated between the 20th and 21st frame predicted by PredNet after being presented with a static gradient image for 20 frames. (B) The height of the bars represents the mean horizontal motion detected by the model, where a positive sign indicates rightward motion. The x-axis labels describe the left and right endpoints of the colour gradient, with the background colour specified in the brackets. Error bars indicate the 99.5% confidence interval of 1000 bootstrap samples of the mean, and red asterisks indicate significant motion. [Figure should be viewed in colour.]*

# 5 Experiment 5: Testing the robustness of results across ten retrained networks

In light of the idea that individual differences exist in the performances and representations of Deep Neural Network instances that are trained with random initial weights[33], we re-ran the above experiments on a newly trained set of PredNet instances with different random initial weights to test the robustness of their responses to the illusory/ components of the illusory stimuli.

## 5.1 Methods:

We trained ten additional PredNet models from different random initial network weights on 820,000 frames of the same 29v dataset. We chose to use twice as many frames of training data as the original pretrained network because Watanabe et al (2018)[21] reported that the prediction of illusory motion only tended to emerge after around 400,000 training frames. For comparability, all other training parameters were the same as those used for the pretrained network: mean squared pixel error loss was minimised using the Adam optimiser with a learning rate of 0.001, reduced by a factor of 0.9 each 100 batches until a minimum of 0.0001, and model weights were updated after each series of 20 frames (i.e. batch size of 1) and all training frames were processed in order (i.e. batch sampling order was not shuffled).

## 5.2 Results:

### 5.2.1 Reproducibility of peripheral drift illusions in PredNet networks

The original PredNet from Watanabe et al (2018) (Figure 6B, leftmost panel) and four of our ten retrained models (Figure 6B, right two rows) predicted counter-clockwise rotational motion when shown the static coloured Rotating Snakes Illusion, consistent with human perception. One model (final panel of second row) predicted clockwise rotational motion, while the remaining five predicted weak or non-rotational motion. No model predicted rotational motion in a closely matched "non-illusion" version of the Rotating Snakes image, in which elements are arranged symmetrically and humans perceive no peripheral drift (Figure 6D). Notably, neither the original PredNet nor any retrained model predicted rotational motion for a grayscale version of the Rotating Snakes image (Figure 6C), despite this inducing the same illusion in human observers[18]. In follow-up tests, we trained ten additional networks on greyscale video footage, and found that they also failed to predict rotational illusory motion (Supplementary Figure 1). Finally, a colour-based variant of the Fraser-Wilcox peripheral drift illusion elicited strong predicted motion from the retrained networks as well as the pretrained network, although there was little consistency across instances of the retrained networks; four networks predicted clockwise rotation and four predicted counter-clockwise rotation (Figure 6E). Overall, there is mixed evidence for perception of peripheral drift illusions in PredNets, even those trained on identical real-world video data.

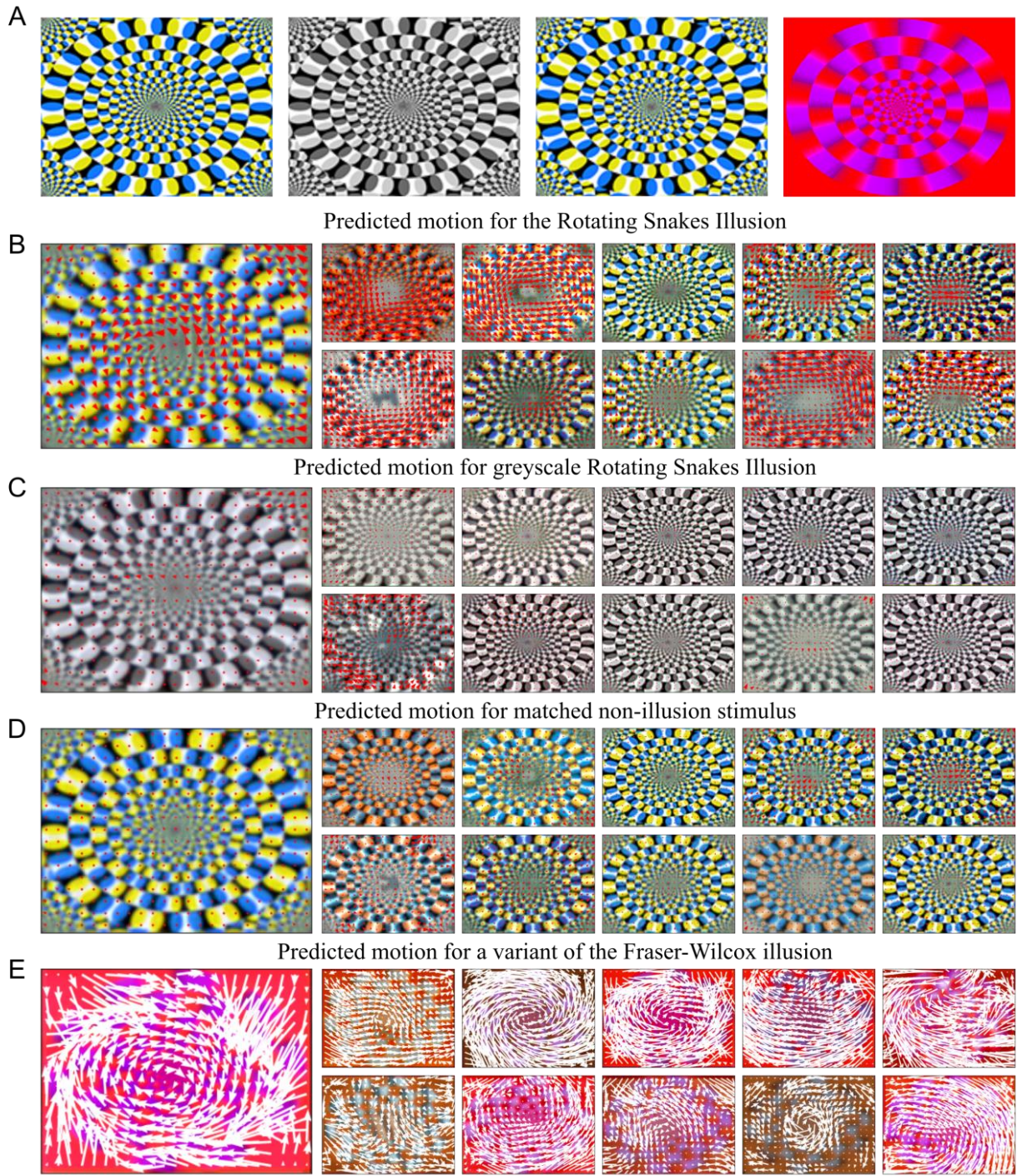

*Figure 6: Motion predicted by ten newly trained PredNets for peripheral drift illusion stimuli.* Optical flow vectors detected between the 21st and 22nd extrapolated frames after networks have received 20 identical static frames of one of the four images shown in (A): (B) the clockwise Rotating Snakes Illusion stimulus, (C) a greyscale version of the same image, (D) a symmetric version of the same pattern, which does not induce illusory motion in human observers, and (E) a colour-based variant of the clockwise Fraser-Wilcox Illusion. All four input images are copyright Akiyoshi Kitaoka. Each section shows in the larger leftmost panel the motion predicted by the original PredNet model pretrained by Watanabe et al. (2018), and in the right two rows that predicted by ten new PredNet models trained from different initial random network weights. [Figure should be viewed in colour.]

**5.2.2 Examining predicted motion for subcomponents of the rotating snakes illusion: Replicating Experiment 2**

The retrained models show high variability in the amount and direction of illusory motion they predict in the bar-pair stimuli of Experiment 2 (Figure 7A). A two-tailed bootstrap confidence interval threshold of 99.5% ($\alpha = 0.05$, Bonferroni multiple comparison corrected for 11 tests, i.e. one per model including the pretrained model) revealed that for the black–blue bar stimulus three out of ten models predict significant horizontal motion rightwards while three of them predict significant motion leftwards. Similarly, for the blue–white bar two models predict significant rightward motion while three models predict significant leftward motion. For the white–yellow bar stimulus, five retrained models predict significant rightward motion and one leftwards, while three models predict significant rightward motion and two models predict significant leftward motion for the yellow–black bar stimulus.

After a similar analysis on the data from grayscale bar stimuli (Figure 7B), we observed that, of the ten retrained models, four predicted significant rightward and three leftward motion on the black–dark gray stimulus. Five models predicted significant rightward motion and one leftward on the dark gray–white stimulus. Three predicted significant rightward motion and three leftward on the white–light gray stimulus. Finally, two and five models predicted significant rightward and leftward motion on the light gray–black stimulus respectively.

The pretrained PredNet's response to the subcomponents of the Rotating Snakes illusion is compatible with human psychophysical results and with the reverse-phi contrast latency theory[18]. However, this is not the case with the retrained models. There are high individual differences in the predicted magnitudes and directions among a set of ten identical PredNet architectures trained on identical data from different random initial weights. There was no apparent consistency in whether a given model predicted leftwards or rightwards motion, nor was there a clear relationship between the motion predicted for these subcomponents and that predicted for the full Rotating Snakes Illusion image (compare Figure 7A and Figure 6B).

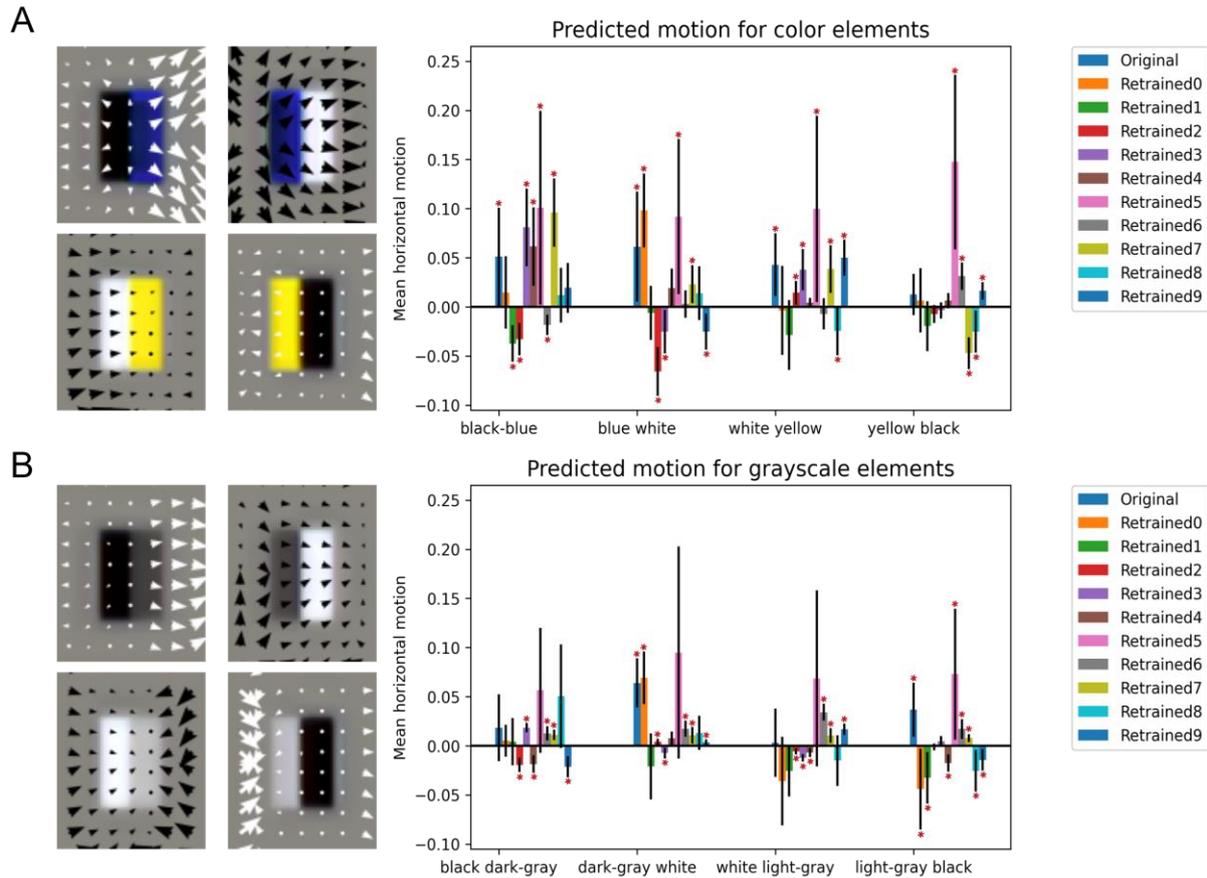

*Figure 7: Predicted motion for subcomponents of the Rotating Snakes Illusion in ten newly-trained PredNets. (A) A colour and (B) a grayscale version of element pairs that make up the rotating snakes illusion are presented to PedNet. Left panels show the optical flow vectors detected in the element pair stimuli by the original network pretrained by Watanabe et al. (2018) (repeated from Figure 3). The arrows indicate the direction and magnitude of optical flow calculated between the 20th and 21st frame predicted by PredNet after being presented with a static element pair image for 20 frames. To the right, barplots show the average direction and magnitude of horizontal motion predicted across all image locations for each stimulus (bar groups) and network (bar colours). The height of the bars represents the mean horizontal motion detected by the model, where a positive sign indicates rightward motion. Means for all the 11 (1 original – pretrained network and 10 retrained networks) PredNet's are plotted for all constituent bar pairs. Error bars indicate the 99.5% confidence interval of 1000 bootstrap samples of the mean (α = 0.05, Bonferroni corrected for 11 comparisons). [Figure should be viewed in colour.]*

### 5.2.3 Effect of contrast and luminance on predicted motion: Replicating Experiment 4

The retrained networks again show variability in the direction and magnitude of predicted motion (Figure 8B). After calculating a two-tailed bootstrap confidence interval of 99.5% on the mean horizontal motion between network-predicted frames for each stimulus, we observed that, of the ten models, three predicted significant rightward motion and five leftward, respectively, for the blue–black gradient on a black background (contrast-dependence predicts leftward motion). Two and five models predicted significant rightward and leftward motion, respectively, for the yellow–white gradient on a white background (contrast-dependence again predicted leftward motion). When the same gradient was presented on a yellow background, seven models predicted significant rightward motion and only one leftward motion (contrast-dependence predicts rightwards motion). Finally, two models predicted significant rightward motion and one leftward for the blue–black gradient on a blue background (contrast-dependence predicts rightwards motion). There is, if anything, a trend towards PredNets predicting motion towards the location of the higher contrast in gradient stimuli, rather than depending on luminance like human perception[37].

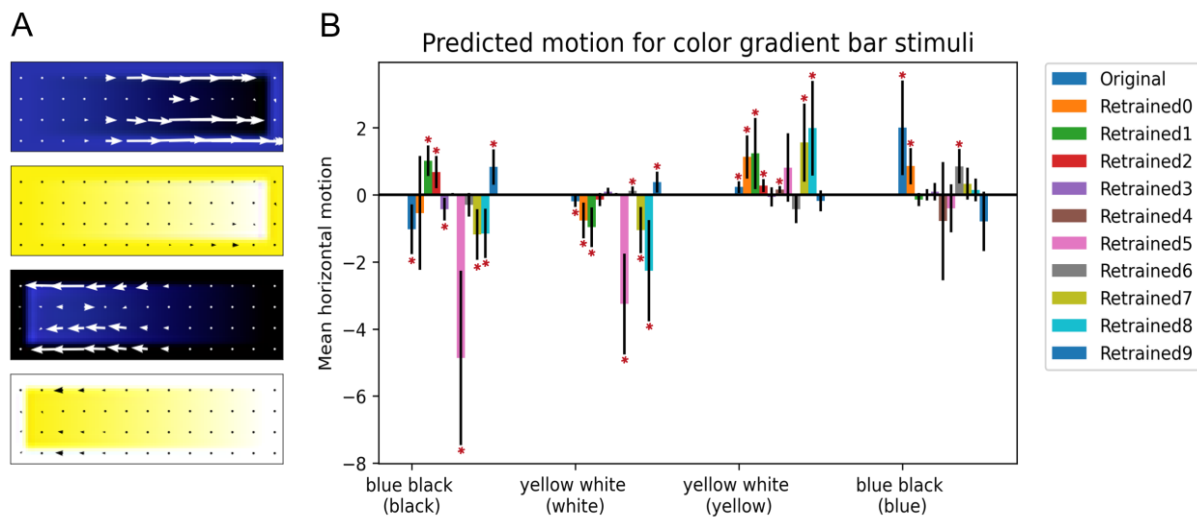

*Figure 8: Predicted motion for gradient bars in ten newly-trained PredNets.* Colour gradient bars similar to the stimuli used by von Grünau et al, 1995 are presented to PredNet. (A) Optical flow vectors detected in the colour gradient bars by the original network pretrained by Watanabe et al. (2018), repeated from Figure 5. The arrows indicate the direction and magnitude of optical flow calculated between the 20th and 21st frame predicted by PredNet after being presented with a static gradient image for 20 frames. (B) The height of the bars represents the mean horizontal motion detected by the model, where a positive sign indicates rightward motion. The x-axis labels describe the left and right endpoints of the colour gradient, with the background colour mentioned in the brackets. Means for all the 11 (one original pretrained network and 10 retrained networks) PredNet's are plotted for all the four gradient stimuli. Error bars indicate the 99.5% confidence interval of 1000 bootstrap samples of the mean ($\alpha = 0.05$, Bonferroni corrected for 11 comparisons). [Figure should be viewed in colour.]

# 6    Discussion:

Deep neural networks (DNNs) have been at the centre of a storm of attention over the past decade, both as engineering solutions to long-standing computer vision challenges, and as potential models of biological vision. DNNs have proved better able to predict brain responses to images than previous image-computable models [46][47][48][49][50], but have had mixed success in reproducing the perceptual idiosyncrasies of human vision[51][52][53][54][55]. Visual illusions involving modal contour completion[56][57], colour and luminance constancy[58], material constancy[59], orientation biases[60], and relative size[61] have been reported in diverse architectures of DNNs, but also failures to reproduce human shape biases[53][62][55], crowding[63], and perception of spatial relations[64][65] and geometric (im-)possibility[66]. Even when a certain network reproduces a certain perceptual phenomenon, our results here demonstrate that its responses to related stimuli may differ from humans', and that even identically-trained networks may behave quite differently.

We replicated and extended the finding that PredNet networks predict motion in the Rotating Snakes Illusion in both a pretrained network (Watanabe et al 2018) as well as networks retrained using the same video footage, but from different random states (Figure 6B). None of the networks predicted coherent illusory motion for a closely-matched non-illusion version of the image (Figure 6D), suggesting that the network doesn't predict motion on all stationary images but only on for particular patterns. We find that PredNet networks are also susceptible to other peripheral drift illusions, specifically a variant of the Fraser-Wilcox illusion[2] (Figure 6E), although the direction of illusory rotation seems less consistent than in humans. PredNet also predicts illusory motion in stimuli that have been reduced to the components of each illusion type such as pairs of coloured or grayscale bars, and coloured gradients, allowing us to probe the nature of the illusory motion predictions more closely. Follow-up studies have also reported illusory motion in PredNet to local patterns in certain natural images[67] and to a wide range of four- and three-colour patterns containing similar luminance relationships to those in the Rotating Snakes Illusion[68][67]. Surprisingly, we found that PredNet fails to predict consistent illusory motion for a grayscale version of the Rotating Snakes Illusion (Figure 6C), which is an effective stimulus for humans[18]. This is not because these images are too far outside of its training domain, since models trained exclusively on grayscale versions of the same footage also do not predict illusory motion for grayscale versions of the illusions (Supplementary Figure 1). It is not clear why the networks not predict motion in grayscale peripheral drift stimuli, but the finding presents a challenge to PredNet's value as a model of human illusory motion.

For the Rotating Snakes Illusion, the pretrained network generally perceives motion in a direction that is consistent with human and macaque perception, for both the full illusion and simplified components[18]. Another study showed recently that the strength of illusory motion predicted by PredNet in simple repeating patterns is related to the number, hue and luminance of their elements[68]. In humans, the greyscale version of the Rotating Snakes illusion undergoes a curious reversal of direction for certain luminance combinations of light-grey and dark-grey elements[45], which is predicted by a motion energy model[19], but since none of our networks predicted motion in a greyscale Rotating Snakes pattern we were not able to test this. The prediction of motion in the same direction for all four subcomponent pairs of the Rotating Snakes Illusion in the pretrained PredNet is consistent with human psychophysical results[18], although the time courses of activation responses within the network do not show the latency differences for different elements of the illusion found in macaque neural data. Predicted motion in simplified gradient stimuli generally tended toward the higher contrast part of a gradient, unlike in humans where illusory motion tends toward the lower luminance part of a gradient[37]. In summary, we find that illusory motion phenomena, even in the pretrained PredNet, do not robustly agree with psychophysical or electrophysiological findings.

One of the most striking aspects of our results is the high degree of variability both within and between networks, even though they were trained on identical data. This is somewhat complementary to the individual differences reported in human perception of peripheral drift illusions, for example in an earlier version of the Rotating Snakes illusion called the "Escalator" illusion designed by Fraser and Wilcox, 1979[2]. In their psychophysics study, 41% of the participants either did not perceive any illusory motion, perceived it in the direction opposite to the majority, or reported it as alternating in direction[2]. Other studies using different variants of the Escalator illusion have reported a similar degree of human individual differences in either the direction or the strength of peripheral drift illusions[1][69][3]. It should be noted that we did not observe any consistent pattern among our 10 retrained PredNets in terms of which stimuli they predicted illusory motion for, or in which direction. A larger sample of networks could be used in the future to investigate whether there is a certain number of distinct "modes" of peripheral drift behaviour.

The reason why PredNet networks often predict motion in peripheral drift illusion stimuli remains ambiguous. One possible interpretation is that the asymmetric patterns in such stimuli reveal differences in the fidelity with which information at different locations in the image is stored by the network as it extrapolates future frames. As future frames are imprecisely extrapolated, the representation of lower-contrast parts of the image may be maintained less well over time (e.g. within the R-unit LSTMs) than that of the higher-contrast parts of the image. Successive extrapolations would therefore fade or blur systematically from the low-to-high contrast end of a gradient, which is detected as a motion signal by the optic flow analysis. However, it is less simple to interpret the Rotating Snakes Illusion in this way, and it does not explain why no networks predicted motion in greyscale versions of illusion images. It has been suggested[70] that PredNet resorts to last frame copying when there is little movement between input frames. Our findings of predicted motion for a wide range of static images, even after twenty frames without movement, demonstrate that this is not strictly true.

We did not optimise our network design or training procedures to yield the highest possible accuracy in predicting future video frames. Rather, we focused on replicating the particular training regime used by Watanabe et al (2018) in order to explore the range of illusion behaviours displayed by different instances of that model. Our retrained networks do appear to be comparable with that previous model: they are on par with or superior to the video-prediction accuracy of the original network, and almost all (eight out of eleven networks) are superior to a baseline of copying the previous frame (Supplementary Figure 2). Several methods to improve prediction accuracy of the network have been proposed which include tweaking hyperparameters, altering the loss function, or revising its architecture[70][71]. However, there was a slight tendency for networks that were better at predicting videos to predict less illusory motion (Supplementary Figure 2), so it doesn't seem likely that optimising for video prediction would enhance the illusory motion effects in this class of network.

The PredNet model is intended to implement the principles of hierarchical error-minimisation in predictive coding, rather than to provide a full simulation of cortical processing, nor was its architecture engineered explicitly to reproduce biological data. Previous work has suggested that the architecture of PredNet is not strictly compatible with Rao and Ballard's predictive coding framework and that the learning ability of the network relies heavily on machine learning hyperparameters like video frame rate, number of layers, and image size. Nevertheless PredNet's recurrent connectivity, hierarchical architecture with increasing receptive field size, and activity minimization through explicit computation of errors in error layers have proved successful in reproducing several neural and perceptual phenomena[56]. Our results suggest that at least PredNet's implementation of predictive processing principles do not necessarily lead to human-like illusory motion, given the large diversity of motion behaviours among our networks. Future work on peripheral drift illusions in DNNs should take our findings as a note of caution, and revisit the question of whether there are architectures and training regimes that do robustly produce such illusions.

The tantalising appeal of DNNs in vision science is that they promise mechanistic understanding of perceptual processes in fully-explicit, biologically-inspired models, and provide powerful new computational methods to describe and generate images in rich feature spaces. However, substantial challenges remain in interpreting the inner workings of DNNs, and in capturing rich comparisons between their perceptual behaviour and that of humans[52][72]. Emerging methods to efficiently discover stimuli that induce "illusions" in DNNs may help us more thoroughly explore similarities and differences between machine and human vision[30][73]. Whether DNN results agree with or diverge from psychophysical data, they provide us with an opportunity to test and refine theories about which computational mechanisms or objective functions are necessary or sufficient to produce human-like vision.

## Acknowledgements

This work was supported by a cluster project grant "The Adaptive Mind", funded by the Excellence Program of the Hessian Ministry of Higher Education, Research, Science, and the Arts (HMWK), Germany, and by a Marsden Fast Start grant from the Royal Society of New Zealand (project MFP-UOA2109). We are grateful to Eiji Watanabe for assistance and code sharing to implement the pretrained model.

## Declaration of interest

The authors declare that they have no competing financial or personal interests.

## Author contributions

**Kirubeswaran**: Conceptualization, Methodology, Software, Formal analysis, Investigation, Writing – Original Draft, Visualization. **Katherine Storrs**: Conceptualization, Methodology, Investigation, Resources, Writing – Review & Editing, Supervision.